\documentclass[12pt]{article}
\usepackage[dvips]{graphicx}

\renewcommand{\bar}[1]{\overline{#1}}
\newcommand{\qu}{{\rm q}}
\newcommand{\qb}{${\rm\bar q}$}
\newcommand{\pvec}{\vec p}
\newcommand{\kvec}{\vec k}
\newcommand{\rvec}{\vec r}
\newcommand{\Rvec}{\vec R}
\newcommand{\ieps}{i\varepsilon}
\newcommand{\pl}{{||}}
\newcommand{\order}[1]{${ O}\left(#1 \right)$}
\newcommand{\eq}[1]{(\ref{#1})}
\newcommand{\half}{{$\frac{1}{2}$}} 
\newcommand{\cd}{\makebox[0.08cm]{$\cdot$}}
\newcommand{\MSbar} {\hbox{$\overline{\hbox{\tiny MS}}$}}
\newcommand{\barMS} {\overline{\rm MS}}

\begin{document}

\begin{flushright}
SLAC-PUB-10161 \\
October 2003
\end{flushright}

\begin{center}
{{\bf\LARGE
New Perspectives for QCD:\\
The Novel Effects of Final-State Interactions and \\
Near-Conformal Effective Couplings\footnote{Work supported by Department of Energy contract
DE-AC03-76SF00515}\\[.2ex]}}

\bigskip
{\it Stanley J. Brodsky \\
Stanford Linear Accelerator Center \\
Stanford University, Stanford, California 94309 \\
 E-mail:  sjbth@slac.stanford.edu }
\end{center}


\begin{center}
{\bf\large Abstract}
\end{center}
\bigskip
The effective QCD charge extracted from  $\tau$ decay is remarkably constant at small
momenta, implying the near-conformal behavior of hadronic interactions at small momentum
transfer. The correspondence of large-$N_C$ supergravity theory in higher-dimensional
anti-de Sitter spaces with gauge theory in physical space-time also has interesting
implications for hadron phenomenology in the conformal limit, such as constituent counting
rules for hard exclusive processes.  The utility of light-front quantization and
light-front Fock wavefunctions for analyzing such phenomena and representing the dynamics
of QCD bound states is reviewed. I also discuss the novel effects of initial- and
final-state interactions in hard QCD inclusive processes, including Bjorken-scaling
single-spin asymmetries and the leading-twist diffractive and shadowing contributions to
deep inelastic lepton-proton scattering.

\eject

\centerline{\large\it  1. Introduction}
\bigskip

The physical masses and inverse sizes of hadrons are characterized by the fundamental mass
scale of QCD: $\Lambda_{QCD} \simeq 200$ MeV. Nevertheless, much QCD phenomenology, such as
Bjorken scaling and dimensional counting rules for hard exclusive processes, can be
understood from the standpoint of a theory without a fundamental scale -- conformal theory.

It is often assumed that color confinement in QCD can be traced to the singular behavior of
the running coupling in the infrared, {\em i.e.} ``infrared slavery."  For example, if
$\alpha_s(q^2) \to \frac{1}{q^2}$ at $q^2 \to 0$, then one-gluon exchange leads to a linear
heavy quark potential at large distances.  However,
theoretical~\cite{vonSmekal:1997is,Zwanziger:2003cf,Howe:2002rb,Aguilar:2002tc,Howe:2003mp,Furui:2003mz}
and phenomenological~\cite{Mattingly:ej,Brodsky:2002nb,Baldicchi:2002qm} evidence is now
accumulating that the QCD coupling becomes
constant~\cite{Parisi:1980jy,Gribov:1998kb,Dokshitzer:1995zt,Shirkov:1997wi} at small
virtuality; {\em i.e.}, $\alpha_s(Q^2)$ develops an infrared fixed point, in contradiction
to the usual assumption of singular growth in the infrared.  If QCD running couplings are
bounded, the integration over the running coupling is finite and renormalon resummations
are  not required. If the QCD coupling becomes scale-invariant in the infrared, then
elements of conformal theory~\cite{Braun:2003rp} become relevant even at relatively small
momentum transfers.

The near-conformal behavior of QCD is the basis for commensurate scale
relations~\cite{Brodsky:1994eh} which relate observables to each other without
renormalization scale or scheme ambiguities~\cite{Brodsky:2000cr}. An important example is
the generalized Crewther relation~\cite{Brodsky:1995tb}.  In this method the effective
charges of observables are related to each other in conformal gauge theory; the effects of
the nonzero QCD $\beta-$ function are then taken into account using the BLM
method~\cite{Brodsky:1982gc} to set the scales of the respective couplings. The conformal
approximation to QCD can also be used as template for QCD
analyses~\cite{Brodsky:1985ve,Brodsky:1984xk} such as the form of the expansion polynomials
for distribution amplitudes~\cite{Braun:1999te}.

Polchinski and Strassler~\cite{Polchinski:2001tt} have derived new results for QCD in the
strong coupling limit using Maldacena's string duality~\cite{Maldacena:1997re},  mapping
features of large $N_C$ supergravity theory  in a higher dimensional  anti-de Sitter space
to conformal gauge theory in 4-dimensional space-time.  This correspondence has many
implications for QCD in the conformal limit, allowing results usually discussed in
perturbation theory such as quark counting
rules~\cite{Brodsky:1973kr,Matveev:ra,Brodsky:1974vy} and $x \to 1$ spectator counting
rules~\cite{Brodsky:1994kg} to be derived at all orders.

\begin{figure}[htbp]
\centering
\includegraphics[height=5in,width=4.9in]{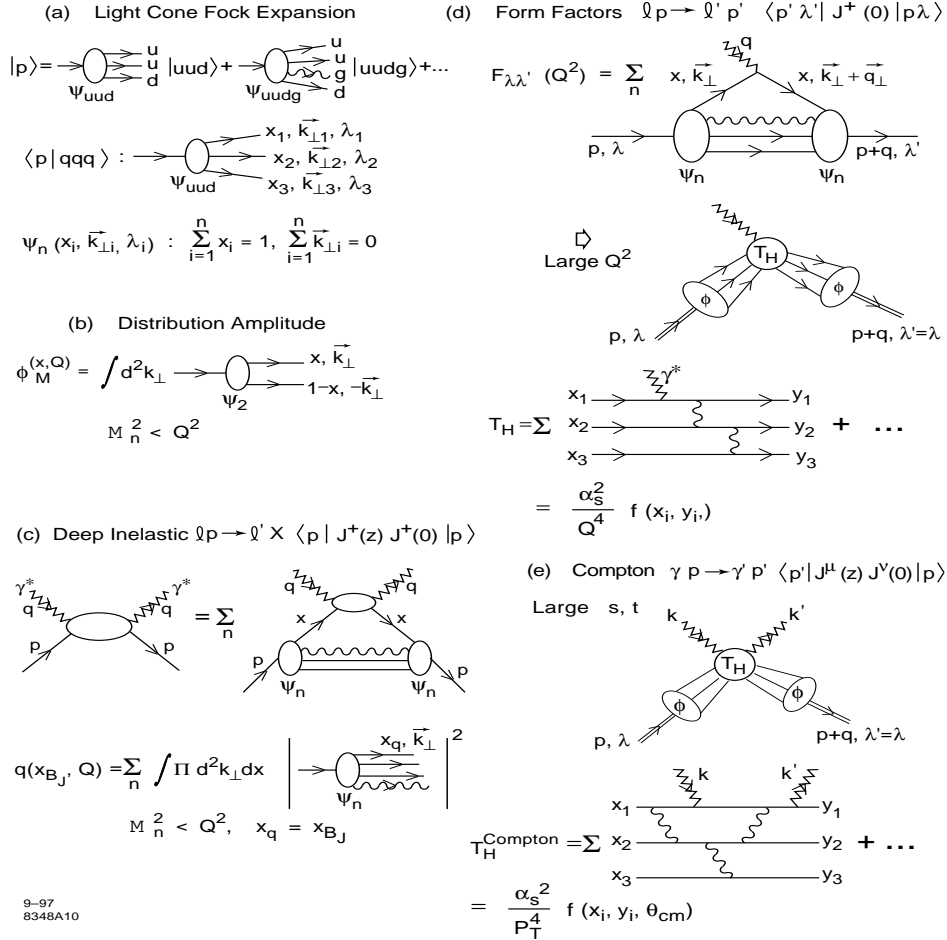}
\caption[*]{Representation of QCD hadronic processes in the
light-cone Fock expansion. (a) The valence $uud$  and $uudg$
contributions to the light-cone Fock expansion for the proton. (b)
The distribution amplitude $\phi(x,Q)$ of a meson expressed as an
integral over its valence light-cone wave function restricted to
$q \bar q$ invariant mass less than $Q$. (c) Representation of
deep inelastic scattering and the quark distributions $q(x,Q)$  as
probabilistic measures of the light-cone Fock wave functions.  The
sum is over the Fock states with invariant mass less than $Q$. (d)
Exact representation of spacelike form factors of the proton in
the light-cone Fock basis.  The sum is over all Fock components.
At large momentum transfer the leading-twist contribution
factorizes as the product of the hard scattering amplitude $T_H$
for the scattering of the valence quarks collinear with the
initial to final direction convoluted with the proton distribution
amplitude.(e) Leading-twist factorization of the Compton amplitude
at large momentum transfer.} \label{Fig:repc1}
\end{figure}

\begin{figure}[htbp]
\centering
 \includegraphics[width=5in]{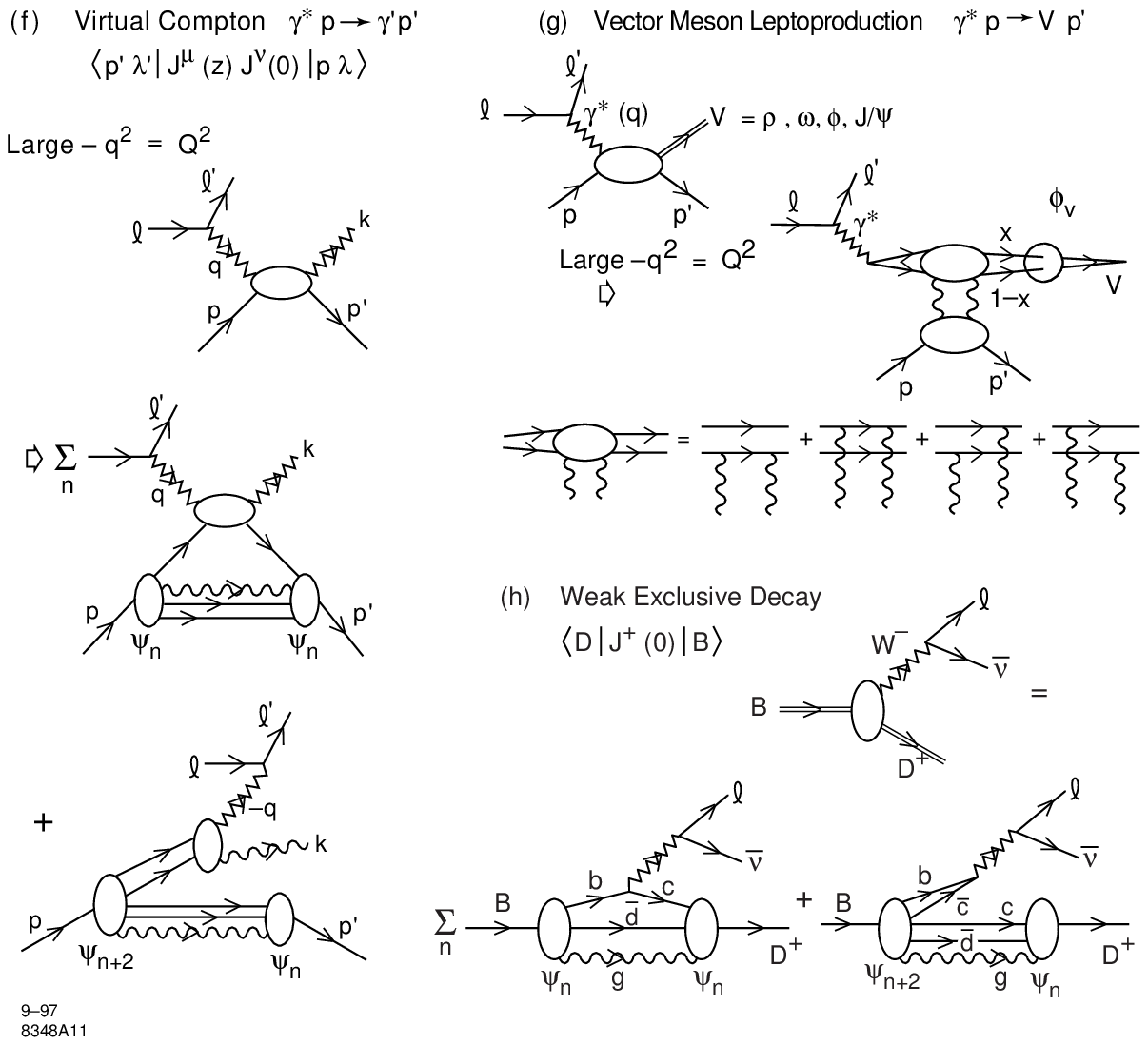}
\caption[*]{(f) Representation of deeply virtual Compton
scattering in the light-cone Fock expansion in the handbag
approximation and at leading twist.  Both diagonal $n  \to n$ and
off-diagonal $n+2 \to n$ contributions are required.  (g)
Diffractive vector meson production at large photon virtuality
$Q^2$ and longitudinal polarization. The high energy behavior
involves two gluons in the $t$ channel coupling to the compact
color dipole structure of the upper vertex.  The bound-state
structure of the vector meson enters through its distribution
amplitude.  (h) Exact representation of the weak semileptonic
decays of heavy hadrons in the light-cone Fock expansion.   Both
diagonal $n  \to n$ and off-diagonal pair annihilation $n+2 \to n$
contributions are required.} \label{Fig:repc2}
\end{figure}

In these lectures, I will also briefly discuss the use of the light-front Fock expansion
for describing the bound-state structure of relativistic composite systems in quantum field
theory. The properties of hadrons are encoded in terms of a set of frame-independent
$n$-particle wave functions~\cite{Brodsky:1997de}. Conformal symmetry and the AdS
correspondence can be used to analyze the asymptotic properties of light-front
wavefunctions, independent of perturbation theory~\cite{deT}.  Light-front quantization in
the doubly-transverse light-cone gauge~\cite{Tomboulis:jn,Srivastava:2000cf}  has a number
of advantages, including explicit unitarity, a physical Fock expansion, exact
representations of current matrix elements, and the decoupling properties needed to prove
factorization theorems in high momentum transfer inclusive and exclusive reactions.  For
example, one can derive exact formulae for the weak decays of the $B$ meson such as $B \to
\ell \bar \nu \pi$~\cite{Brodsky:1998hn}  and the deeply virtual Compton amplitude in the
handbag approximation~\cite{Brodsky:2000xy,Diehl:2000xz}.  Applications include two-photon
exclusive reactions, and diffractive dissociation into jets.  The universal light-front
wave functions and distribution amplitudes control hard exclusive processes such as form
factors, deeply virtual Compton scattering, high momentum transfer photoproduction, and
two-photon processes. The utility of light-front wave functions for the computation of
various exclusive and inclusive processes is illustrated in Figs.~\ref{Fig:repc1} and
\ref{Fig:repc2}.

A new understanding of the role of final-state interactions in deep inelastic scattering
has recently emerged~\cite{Brodsky:2002ue}.  The final-state interactions from gluon
exchange between the outgoing quark and the target spectator system lead to single-spin
asymmetries in semi-inclusive deep inelastic lepton-proton scattering at leading twist in
perturbative QCD; {\em i.e.}, the rescattering corrections of the struck quark with the
target spectators are not power-law suppressed at large photon virtuality $Q^2$ at fixed
$x_{bj}$~\cite{Brodsky:2002cx}  The final-state interaction from gluon exchange occurring
immediately after the interaction of the current also produces a leading-twist diffractive
component to deep inelastic scattering $\ell p \to \ell^\prime p^\prime X$ corresponding to
color-singlet exchange with the target system; this in turn produces shadowing and
anti-shadowing of the nuclear structure functions~\cite{Brodsky:2002ue,Brodsky:1989qz}.  In
addition, one can show that the pomeron structure function derived from diffractive DIS has
the same form as the quark contribution of the gluon structure function~\cite{BHIE}.  The
final-state interactions occur at a light-cone time $\Delta\tau \simeq 1/\nu$ after the
virtual photon interacts with the struck quark, producing a nontrivial phase.  Thus none of
the above phenomena is contained in the target light-front wave functions computed in
isolation.   In particular, the shadowing of nuclear structure functions is due to
destructive interference effects from leading-twist diffraction of the virtual photon,
physics not included in the nuclear light-cone wave functions.  Thus the structure
functions measured in deep inelastic lepton scattering are affected by final-state
rescattering, modifying their connection to light-front probability distributions.  Some of
these results can be understood by augmenting the light-front wave functions with a gauge
link, but with a gauge potential created by an external field created by the virtual photon
$q \bar q$ pair current~\cite{Belitsky:2002sm}.  The gauge link is also process
dependent~\cite{Collins:2002kn}, so the resulting augmented LFWFs are not universal.

\bigskip

 \centerline{\large\it  2. The Behavior of the Effective QCD
Coupling $\alpha_{\tau}(s)$}
\bigskip

One can define the fundamental coupling of QCD from virtually any
physical observable~\cite{Grunberg:1980ja,Grunberg:1982fw}.  Such
couplings, called effective charges, are all-order resummations of
perturbation theory, so they  correspond to the complete theory of
QCD; it is thus guaranteed that they are analytic and
non-singular.  For example, it has been shown that unlike the
$\barMS$ coupling, a physical coupling is analytic across quark
flavor thresholds~\cite{Brodsky:1998mf,Brodsky:1999fr}.
Furthermore, a physical coupling must stay finite in the infrared
when the momentum scale goes to zero.  In turn, this means that
integrals over the running coupling are well defined for physical
couplings.  Once such a physical coupling $\alpha_{\rm phys}(k^2)$
is chosen, other physical quantities can be expressed as
expansions in $\alpha_{\rm phys}$ by eliminating the $\barMS$
coupling which now becomes only an
intermediary~\cite{Brodsky:1994eh}.  In such a procedure there are
in principle no further renormalization scale ($\mu$) or scheme
ambiguities.  The physical couplings satisfy the standard
renormalization group equation for its logarithmic derivative,
${{\rm d}\alpha_{\rm phys}/{\rm d}\ln k^2} = \widehat{\beta}_{\rm
phys}[\alpha_{\rm phys}(k^2)]$, where the first two terms in the
perturbative expansion of the Gell-Mann Low function
$\widehat{\beta}_{\rm phys}$ are scheme-independent at leading
twist, whereas the higher order terms have to be calculated for
each observable separately using perturbation theory.

In a recent paper, Menke, Merino, and
Rathsman~\cite{Brodsky:2002nb} and I have  presented a definition
of a physical coupling for QCD which has a direct relation to high
precision measurements of the hadronic decay channels of the
$\tau^- \to \nu_\tau {\rm h}^-$.  Let $R_{\tau}$ be the ratio of
the hadronic decay rate to the leptonic one.  Then $R_{\tau}\equiv
R_{\tau}^0\left[1+{\alpha_\tau \over \pi}\right]$, where
$R_{\tau}^0$ is the zeroth order QCD prediction, defines the
effective charge $\alpha_\tau$.  The data for $\tau$ decays is
well-understood channel by channel, thus allowing the calculation
of the hadronic decay rate and the effective charge as a function
of the $\tau$ mass below the physical mass.  The vector and
axial-vector decay modes which can be studied separately.

Using an analysis of the tau data from the OPAL
collaboration~\cite{Ackerstaff:1998yj}, we have found that the
experimental value of the coupling
$\alpha_{\tau}(s)=0.621\pm0.008$ at $s = m^2_\tau$ corresponds to
a value of $\alpha_{\MSbar}(M^2_Z) = (0.117$-$0.122) \pm 0.002$,
where the range corresponds to three different perturbative
methods used in analyzing the data.  This result is, at least for
the fixed order and renormalon resummation methods, in good
agreement with the world average $\alpha_{\MSbar}(M^2_Z) = 0.117
\pm 0.002$.  However, from the figure we also see that the
effective charge only reaches $\alpha_{\tau}(s) \sim 0.9 \pm 0.1$
at $s=1\,{\rm GeV}^2$, and it even stays within the same range
down to $s\sim0.5\,{\rm GeV}^2$.  This result is in good agreement
with the estimate of Mattingly and Stevenson~\cite{Mattingly:ej}
for the effective coupling $\alpha_R(s) \sim 0.85 $ for $\sqrt s <
0.3\,{\rm GeV}$ determined from ${\rm e}^+{\rm e}^-$ annihilation,
especially if one takes into account the perturbative commensurate
scale relation, $\alpha_{\tau}(m_{\tau^\prime}^2)= \alpha_R(s^*)$
where, for $\alpha_R=0.85$, we have $s^* \simeq
0.10\,m_{\tau^\prime}^2.$ This behavior is not consistent with the
coupling having a Landau pole, but rather shows that the physical
coupling is close to constant at low scales, suggesting that
physical QCD couplings are effectively constant or ``frozen" at
low scales.

\begin{figure}[htb]
\centering
 \includegraphics[width=3.5in]{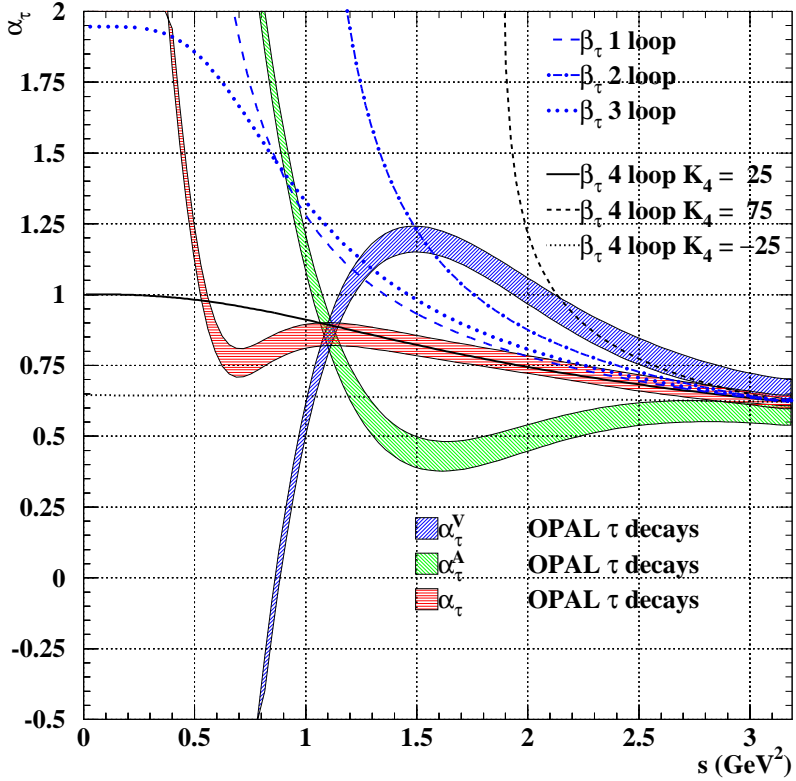}\vspace{.52in}
\caption[*]{ The effective charge $\alpha_{\tau}$ for non-strange
hadronic decays of a hypothetical $\mathit{\tau}$ lepton with
$\mathit{m_{\tau'}^2 = s}$ compared to solutions of the fixed
order evolution equation for $\alpha_{\tau}$ at two-, three-, and
four-loop order.  The error bands include statistical and
systematic errors.} \label{fig:fopt_comp}
\end{figure}

Figure~\ref{fig:fopt_comp} shows a comparison of the
experimentally determined effective charge $\alpha_{\tau}(s)$ with
solutions to the evolution equation for $\alpha_{\tau}$ at two-,
three-, and four-loop order normalized at $m_\tau$.  At three
loops the behavior of the perturbative solution drastically
changes, and instead of diverging, it freezes to a value
$\alpha_{\tau}\simeq 2$ in the infrared.  The reason for this
fundamental change is, the negative sign of $\beta_{\tau,2}$. This
result is not perturbatively stable since the evolution of the
coupling is governed by the highest order term.  This is
illustrated by the widely different results obtained for three
different values of the unknown four loop term $\beta_{\tau,3}$
which are also shown\footnote{The values of $\beta_{\tau,3}$ used
are obtained from the estimate of the four loop term in the
perturbative series of $R_\tau$, $K_4^{\overline{\rm MS}} = 25\pm
50$~\cite{LeDiberder:1992fr}.} It is interesting to note that the
central four-loop solution is in good agreement with the data all
the way down to $s\simeq1\,{\rm GeV}^2$.

The results for $\alpha_{\tau}$ resemble the behavior of the
one-loop ``time-like" effective
coupling~\cite{Beneke:1994qe,Ball:1995ni,Dokshitzer:1995qm}
\begin{equation}\label{eq:alphaeff}
\alpha_{\rm eff}(s)=\frac{4\pi}{\beta_0} \left\{\frac{1}{2} -
\frac{1}{\pi}\arctan\left[\frac{1}{\pi}\ln\frac{s}{\Lambda^2}\right]\right\}
\end{equation}
which is finite in the
infrared and freezes to the value $\alpha_{\rm
eff}(s)={4\pi}/{\beta_0}$ as $s\to 0$.  It is  instructive to
expand the ``time-like" effective coupling for large $s$,
\begin{eqnarray*}
\alpha_{\rm eff}(s) &=&\frac{4\pi}{\beta_0\ln\left(s/\Lambda^2\right)}
\left\{1 -\frac{1}{3}\frac{\pi^2}{\ln^2\left(s/\Lambda^2\right)}
+\frac{1}{5}\frac{\pi^4}{\ln^4\left(s/\Lambda^2\right)} +\ldots
\right\}\\ &=&\alpha_{\rm s}(s)\left\{1
-\frac{\pi^2\beta_0^2}{3}\left(\frac{\alpha_{\rm s}(s)}{4\pi}\right)^2
+\frac{\pi^4\beta_0^4}{5}\left(\frac{\alpha_{\rm s}(s)}{4\pi}\right)^4
+\ldots \right\}.
\end{eqnarray*}
This shows that the ``time-like" effective coupling is a resummation
of $(\pi^2\beta_0^2\alpha_{\rm s}^2)^n$-corrections to the usual running
couplings.  The finite coupling $\alpha_{\rm eff}$ given in
Eq.~(\ref{eq:alphaeff}) obeys standard PQCD evolution at LO.  Thus one
can have a solution for the perturbative running of the QCD coupling
which obeys asymptotic freedom but does not have a Landau singularity.

Recently it has been argued that $\alpha_R(s)$ freezes perturbatively to all
orders~\cite{Howe:2002rb}.  This should also be true perturbatively for $\alpha_{\tau}(s)$.
In fact since all observables are related by commensurate scale relations, they all should
have an IR fixed point~\cite{Howe:2003mp}.  This result is also consistent with
Dyson-Schwinger equation studies of the physical gluon propagator in Landau
gauge~\cite{vonSmekal:1997is}. In contrast, Cucchieri and Zwanziger
~\cite{Cucchieri:2002su,Zwanziger:2002sh,Greensite:2003xf,Zwanziger:2003cf,Swanson:2003sf}
have shown that the QCD coupling defined from the $D_{44}$ term in the Coulomb gauge
propagator in quenched lattice gauge theory exhibits a confining $1\over k^2$ behavior. For
a discussion on how to reconcile these disparate results, see
ref.~\cite{Szczepaniak:2003ve}.

The near constancy of the effective QCD coupling at small scales
helps explain the empirical success of dimensional counting rules
for the power law fall-off of form factors and fixed angle
scaling.  As shown in Refs.~\cite{Brodsky:1997dh,Melic:2001wb},
one can calculate the hard scattering amplitude $T_H$ for such
processes~\cite{Lepage:1980fj} without scale ambiguity in terms of
the effective charge $\alpha_\tau$ or $\alpha_R$ using
commensurate scale relations.  The effective coupling is evaluated
in the regime where the coupling is approximately constant, in
contrast to the rapidly varying behavior from powers of
$\alpha_{\rm s}$ predicted by perturbation theory (the universal
two-loop coupling).  For example, the nucleon form factors are
proportional at leading order to two powers of $\alpha_{\rm s}$
evaluated at low scales in addition to two powers of $1/q^2$; The
pion photoproduction amplitude at fixed angles is proportional at
leading order to three powers of the QCD coupling.  The essential
variation from leading-twist counting-rule behavior then only
arises from the anomalous dimensions of the hadron distribution
amplitudes.

The magnitude of the effective charge~\cite{Brodsky:1997dh}
$\alpha^{\rm exclusive}_s(Q^2) =$ \hfill\break $ {F_\pi(Q^2)/ 4\pi
Q^2 F^2_{\gamma \pi^0}(Q^2)}$ for exclusive amplitudes is
connected to $\alpha_\tau$ by a commensurate scale relation.  Its
magnitude: $\alpha^{\rm exclusive}_s(Q^2) \sim 0.8$ at small
$Q^2,$  is sufficiently large as to explain the observed magnitude
of exclusive amplitudes such as the pion form factor using the
asymptotic distribution amplitude.

\bigskip
\centerline{\large\it  3.  Connections between QCD and Conformal Field
Theory}
\bigskip

As shown by Maldacena~\cite{Maldacena:1997re}, there is a
remarkable correspondence between large $N_C$ supergravity theory
in a higher dimensional  anti-de Sitter space  and supersymmetric
QCD in 4-dimensional space-time.  Recently, Polchinski and
Strassler~\cite{Polchinski:2001tt} have shown that one can use the
Maldacena correspondence to compute the leading power-law falloff
of exclusive processes such as high-energy fixed-angle scattering
of gluonium-gluonium scattering in supersymmetric QCD.  The
power-law fall-off in the gauge theory  reflects the warped
geometry of the anti-de Sitter space in the dual theory.  The
resulting predictions for hadron physics
coincide~\cite{Polchinski:2001tt,Brower:2002er, Andreev:2002aw}
with QCD dimensional counting
rules:~\cite{Brodsky:1973kr,Matveev:ra,Brodsky:1974vy}
\begin{equation}{d\sigma_{H_1 H_2 \to H_3 H_4}\over dt} ={ F(t/s)\over
s^{n-2}}\end{equation} where $n$ is the sum of the minimal number
of interpolating fields.  (For a recent review of hard fixed
$\theta_{CM}$  angle exclusive processes in QCD see
~\cite{Brodsky:2002st}.) As shown by Brower and
Tan~\cite{Brower:2002er}, the non-conformal dimensional scale
which appears in the QCD analysis is set by the string constant,
the  slope of the primary Regge trajectory
$\Lambda^2=\alpha^\prime_R(0)$ of the supergravity theory.
Polchinski and Strassler~\cite{Polchinski:2001tt} have also
derived counting rules for deep inelastic structure functions at
$x \to 1$ in agreement with perturbative QCD
predictions~\cite{Brodsky:1994kg} as well as Bloom-Gilman
exclusive-inclusive duality.

As discussed in the previous section, the QCD running coupling
$\alpha_s(m^2_\tau)$ derived from hadronic $\tau$ decays is
observed to be remarkably flat as a function of the $\tau$ mass,
suggesting that QCD itself has  an infrared fixed point and nearly
conformal behavior at small virtuality.

There are other features of the superstring derivation which are of
interest for QCD phenomenology:

\begin{enumerate}

\item The supergravity analysis is based on an extension of
classical gravity theory in higher dimensions and is
nonperturbative.  Thus the usual analyses of exclusive processes,
which were derived in perturbation theory can be extended by the
Maldacena correspondence to all orders.  An interesting point is
that the hard scattering amplitudes which are normally or order
$\alpha_s^p$ in PQCD appear as order $\alpha_s^{p/2}$ in the
supergravity predictions.  This can be understood as an all-orders
resummation of the effective
potential~\cite{Maldacena:1997re,Rey:1998ik}.

\item The superstring theory results are derived in the limit of a
large $N_C$~\cite{'tHooft:1973jz}.  For gluon-gluon scattering,
the amplitude scales as ${1}/{N_C^2}$.  Frampton has shown how to
extend the analysis to the fundamental
representation~\cite{Frampton:2003kn}.  For color-singlet bound
states of quarks, the amplitude scales as ${1}/{N_C}$.  This large
$N_C$-counting in fact corresponds to the quark interchange
mechanism~\cite{Gunion:1973ex}.  For example, for $K^+ p \to K^+
p$ scattering, the $u$-quark exchange amplitude scales
approximately as $\frac{1}{u}$\ $\frac{1}{t^2},$ which agrees
remarkably well with the measured large $\theta_{CM}$ dependence
of the $K^+ p$ differential cross section~\cite{Sivers:1975dg}.
This implies that the nonsinglet Reggeon trajectory asymptotes to
a negative integer~\cite{Blankenbecler:1973kt}, in this case,
$\lim_{-t \to \infty}\alpha_R(t) \to -1.$

\item  Pinch contributions corresponding to the independent scattering mechanism of
Landshoff~\cite{Landshoff:ew} are absent in the superstring derivation.  This can be
understood by the fact that amplitudes based on gluon exchange between color-singlet
hadrons is suppressed at large $N_C$~\cite{deT}.  Furthermore, the independent scattering
amplitudes are suppressed by Sudakov form factors which fall faster than any power in a
theory with a fixed-point coupling such as conformal QCD~\cite{Brodsky:1974vy,Duncan:ny}.

\item The leading-twist results for hard exclusive processes
correspond to the suppression of hadron wave functions with
non-zero orbital angular momentum, which is the  principle
underlying the selection rules corresponding to hadron helicity
conservation~\cite{Brodsky:1981kj}.  The suppression can be
understood as follows: the LF wave function with nonzero  angular
momentum in the constituent rest frame $\sum \vec k_i = 0$ can be
determined by iterating the one gluon exchange kernel.  They then
have the structure~\cite{Ji:bw,Karmanov} \begin{equation}
\psi_{L_z=1}= {\vec S \cdot \widehat n \times \vec k_\perp\over
D(k^2_\perp,x)} \psi_{L_z=0}\end{equation} or
\begin{equation}\psi_{L_z=1}={ \widehat \epsilon \cdot \widehat n \times \vec
k_\perp\over D(k^2_\perp,x) }\psi_{L_z=0}\end{equation} where the
light-front energy denominator $D(k^2_\perp,x) \sim k^2_\perp$ at
high transverse momentum, $\widehat n$ is the light-front
quantization direction, and $\widehat \epsilon$ is a spin-one
polarization vector. This leads to the $\Lambda/Q$ suppression of
spin-flip amplitudes in QCD.  For example, such wave functions
lead to the large momentum transfer prediction $A_{LL}\sim 1/3$
for $pp \rightarrow pp$ elastic scattering~\cite{deT} at large
angles and momentum transfer and the asymptotic prediction
$F_2(t)/ F_1(t) \propto t^{-2}$ modulo powers of $\log
t$~\cite{Belitsky:2002kj}.

\end{enumerate}

\bigskip
\centerline{\large\it  4. Light-Front Wave Functions and Angular Momentum}
\bigskip

The concept of a wave function of a hadron as a composite of
relativistic quarks and gluons is naturally formulated in terms of
the light-front Fock expansion at fixed light-front time, $\tau=x
\cd \omega$.  The four-vector $\omega$, with $\omega^2 = 0$,
determines the orientation of the light-front plane; the freedom
to choose $\omega$ provides an explicitly covariant formulation of
light-front quantization~\cite{cdkm}. The light-front wave
functions (LFWFs) $\psi_n(x_i,k_{\perp_i},\lambda_i)$, with
$x_i={k_i \cd \omega\over P\cd \omega}$, $\sum^n_{i=1} x_i=1, $
$\sum^n_{i=1}k_{\perp_i}=0_\perp$, are the coefficient functions
for $n$ partons in the Fock expansion, providing a general
frame-independent representation of the hadron state.  Matrix
elements of local operators such as spacelike proton form factors
can be computed simply from the overlap integrals of light front
wave functions in analogy to nonrelativistic Schr\"odinger theory.
In principle, one can solve for the LFWFs directly from the
fundamental theory using methods such as discretized light-front
quantization, the transverse lattice, lattice gauge theory
moments, or Bethe--Salpeter techniques.  The determination of the
hadron LFWFs from phenomenological constraints and from QCD itself
is a central goal of hadron and nuclear physics.  Reviews of
nonperturbative light-front methods may be found in
Refs.~\cite{Brodsky:1997de,cdkm,Dalley:ug}.

One of the central issues in the analysis of fundamental hadron
structure is the presence of non-zero orbital angular momentum in
the bound-state wave functions.  The evidence for a ``spin crisis"
in the Ellis-Jaffe sum rule signals a significant orbital
contribution in the proton wave
function~\cite{Jaffe:1989jz,Ji:2002qa}.  The Pauli form factor of
nucleons is computed from the overlap of LFWFs differing by one
unit of orbital angular momentum $\Delta L_z= \pm 1$.  Thus the
fact that the anomalous moment of the proton is non-zero requires
nonzero orbital angular momentum in the proton
wavefunction~\cite{BD80}.  In the light-front method, orbital
angular momentum is treated explicitly; it includes the orbital
contributions induced by relativistic effects, such as the
spin-orbit effects normally associated with the conventional Dirac
spinors.

A number of new non-perturbative methods for determining light-front wave functions have
been developed including discretized light-cone quantization using Pauli-Villars
regularization, supersymmetry, and the transverse lattice.  One can also project the known
solutions of the Bethe-Salpeter equation to equal light-front time, thus producing hadronic
light-front Fock wave functions.  A potentially important method is to construct the $q\bar
q$ Green's function using light-front Hamiltonian theory, with DLCQ boundary conditions and
Lippmann-Schwinger resummation.  The zeros of the resulting resolvent projected on states
of specific angular momentum $J_z$ can then generate the meson spectrum and their
light-front Fock wavefunctions.  For a recent review of light-front methods and references,
see Ref.~\cite{Brodsky:2003gk}.

Diffractive multi-jet production in heavy nuclei provides a novel
way to measure the shape of light-front Fock state wave functions
and test color transparency~\cite{Brodsky:xz}. For example,
consider the reaction~\cite{Bertsch:1981py,Frankfurt:1999tq} $\pi
A \rightarrow {\rm Jet}_1 + {\rm Jet}_2 + A^\prime$ at high energy
where the nucleus $A^\prime$ is left intact in its ground state.
The transverse momenta of the jets balance so that $ \vec k_{\perp
i} + \vec k_{\perp 2} = \vec q_\perp < {R^{-1}}_A \ . $ The
light-cone longitudinal momentum fractions also need to add to
$x_1+x_2 \sim 1$ so that $\Delta p_L < R^{-1}_A$.  The process can
then occur coherently in the nucleus.  Because of color
transparency, the valence wave function of the pion with small
impact separation, will penetrate the nucleus with minimal
interactions, diffracting into jet pairs~\cite{Bertsch:1981py}.
The $x_1=x$, $x_2=1-x$ dependence of the di-jet distributions will
thus reflect the shape of the pion valence light-cone wave
function in $x$; similarly, the $\vec k_{\perp 1}- \vec k_{\perp
2}$ relative transverse momenta of the jets gives key information
on the derivative of the underlying shape of the valence pion
wavefunction~\cite{Frankfurt:1999tq,Nikolaev:2000sh}. The
diffractive nuclear amplitude extrapolated to $t = 0$ should be
linear in nuclear number $A$ if color transparency is correct. The
integrated diffractive rate should then scale as $A^2/R^2_A \sim
A^{4/3}$ as verified by E791 for 500 GeV incident pions on nuclear
targets~\cite{Aitala:2000hc}.  The measured momentum fraction
distribution of the jets~\cite{Aitala:2000hb} is consistent with
the shape of the pion asymptotic distribution amplitude,
$\phi^{\rm asympt}_\pi (x) = \sqrt 3 f_\pi x(1-x)$. Data from
CLEO~\cite{Gronberg:1998fj} for the $\gamma \gamma^* \rightarrow
\pi^0$ transition form factor also favor a form for the pion
distribution amplitude close to the asymptotic solution to its
perturbative QCD evolution
equation~\cite{Lepage:1979zb,Efremov:1978rn,Lepage:1980fj}.

In recent work, Dae Sung Hwang, John Hiller, Volodya
Karmanov~\cite{Karmanov}, and I  have studied the analytic
structure of LFWFs using the explicitly Lorentz-invariant
formulation of the front form.  Eigensolutions of the
Bethe-Salpeter equation have specific angular momentum as
specified by the Pauli-Lubanski vector.  The corresponding LFWF
for an $n$-particle Fock state evaluated at equal light-front time
$\tau = \omega\cdot x$ can be obtained by integrating the
Bethe-Salpeter solutions over the corresponding relative
light-front energies.  The resulting LFWFs $\psi^I_n(x_i, k_{\perp
i})$ are functions of the light-cone momentum fractions $x_i=
{k_i\cdot \omega / p \cdot \omega}$ and the invariant mass squared
of the constituents $M_0^2= (\sum^n_{i=1} k_i^\mu)^2 =\sum_{i
=1}^n \big [{k^2_\perp + m^2\over x}\big]_i$ and the light-cone
momentum fractions $x_i= {k\cdot \omega / p \cdot \omega}$ each
multiplying spin-vector and polarization tensor invariants which
can involve $\omega^\mu.$  The resulting LFWFs for bound states
are eigenstates of the Karmanov--Smirnov kinematic angular
momentum operator~\cite{ks92}. Thus LFWFs satisfy all Lorentz
symmetries of the front form, including boost invariance, and they
are proper eigenstates of angular momentum.   Although LFWFs
depend on the choice of the light-front quantization direction,
all observables such as matrix elements of local current
operators, form factors, and cross sections are light-front
invariants -- they must be independent of $\omega_\mu.$

\begin{figure}[ht!b]
\centering
 \includegraphics[width=5in]{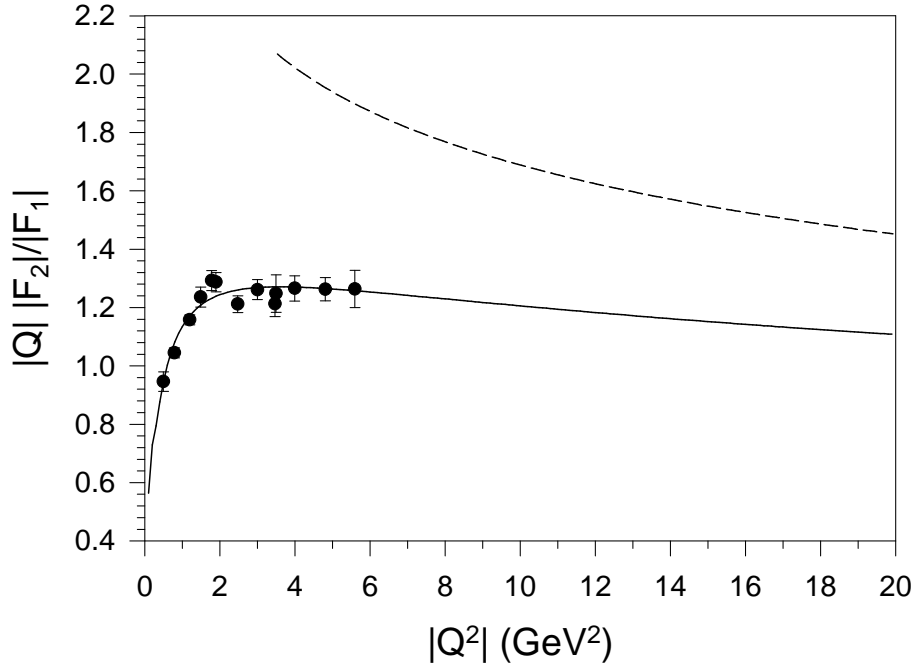}
\caption[*]{Perturbative QCD motivated fit to the Jefferson
Laboratory polarization transfer
data~\cite{Jones:1999rz,Gayou:2001qd}.  The parametrization is
given in Eq.~(\ref{eq:F2F1}) of the text.  The dashed line shows
the predicted form for timelike $q^2=-Q^2$. For a discussion on
the validity of continuing spacelike form factors to the timelike
region, see~\cite{Geshkenbein74} } \label{fig:QF2F1}
\end{figure}

The dependence of the LFWFs on the square of the invariant mass
implies that hadron form factors computed from the overlap
integrals of LFWFS are analytic functions of $Q^2.$  In
particular, the  general form of the LFWFs for baryons in QCD
leads to a ratio of form factors $F_2(Q^2)/F_1(Q^2)$ which behaves
asymptotically as an inverse power of $Q^2$ modulo logarithms, in
agreement with the PQCD analysis of Belitsky, Ji, and
Yuan~\cite{Belitsky:2002kj} as well as with form factor ratios
obtained using the nonperturbative solutions to the Wick--Cutkosky
model found by Karmanov and Smirnov~\cite{ks92}.  The detailed
analysis of baryon form factors at large $Q^2$ based on
perturbative QCD predicts the asymptotic behavior ${Q^2
F_2(Q^2)/F_1(Q^2)} \sim \log^{2+8/(9\beta)}{(Q^2/\Lambda^2)}$,
where $\beta=11-2n_f/3$~\cite{Belitsky:2002kj}.  This asymptotic
logarithmic form can be generalized to include the correct $Q^2=0$
limit and the cut at the two-pion threshold in the timelike
region.  Such a parametrization is
\begin{equation} \label{eq:F2F1}
F_2/F_1 = \kappa_p\frac{1+(Q^2/C_1)^2 \log^{b+2}(1+Q^2/4m_\pi^2)}
               {1+(Q^2/C_2)^3 \log^b(1+Q^2/4m_\pi^2)},
\end{equation}
where for simplicity we have ignored the small factor $8/9\beta$,
as in Ref.~\cite{Belitsky:2002kj}.  For the large-$Q^2$ region of
the available data, this already reduces to the asymptotic form
\begin{equation}
F_2/F_1 = \kappa_p\frac{C_2^3}{C_1^2} \frac{\log^2(Q^2/4m_\pi^2)}{Q^2}.
\end{equation}
The values of $C_1$, $C_2$ and b are not tightly constrained,
except for the combination $C_2^3/C_1^2$. A fit to the JLab data
yields $C_1=0.791$ GeV$^2$, $C_2=0.380$ GeV$^2$, and $b=5.102$.
Thus, as shown in Fig.~\ref{fig:QF2F1}, one can fit the form
factor ratio over the entire measured range with an analytic form
compatible with the predicted perturbative QCD asymptotic
behavior.  The form for spacelike form factors can be analytically
continued to the time-like regime; in particular, one can test the
predicted relative phase of the proton time-like form factors by
measuring the single-spin asymmetry of the produced proton
polarization normal to its production plane in $e^+ e^- \to p \bar
p$~\cite{BCHH}.

Recent advances in the calculation of hard exclusive amplitude at
higher order are discussed by  Duplancic and Nizic in
Ref.~\cite{Duplancic:2003jm} Hadronic exclusive processes are
closely related to exclusive hadronic $B$ decays, processes which
are essential for determining the CKM phases and the physics of
$CP$
violation~\cite{Beneke:2000ry,Keum:2000wi,Szczepaniak:1990dt,Brodsky:2001jw}%
, such as $B \to K \pi$, $B \to \ell \nu \pi$, and $B \to K  p \bar p$ \cite{Chua:2002wn}.
Recently Fred Goldhaber, Jungil Lee and I~\cite{Brodsky:2003hv} have shown how one can
compute the exclusive production of a glueball in association with a charmonium state in
$e^+ e^-$ annihilation.  Since the subprocesses $\gamma^* \to (c\bar c) (c \bar c)$ and
$\gamma^* \to (c \bar c) (g g)$ are of the same nominal order in perturbative QCD, it is
possible that some portion of the anomalously large signal observed by
Belle~\cite{Abe:2002rb} in $e^+ e^- \to J/\psi X$ may actually be due to the production of
charmonium-glueball $J/\psi \mathcal{G}_J$ pairs.

\bigskip

\centerline{\large\it  5. Effects of Final-State Interactions in QCD}
\bigskip

\begin{figure}[htbp]
\begin{center}
\includegraphics[width=5in]{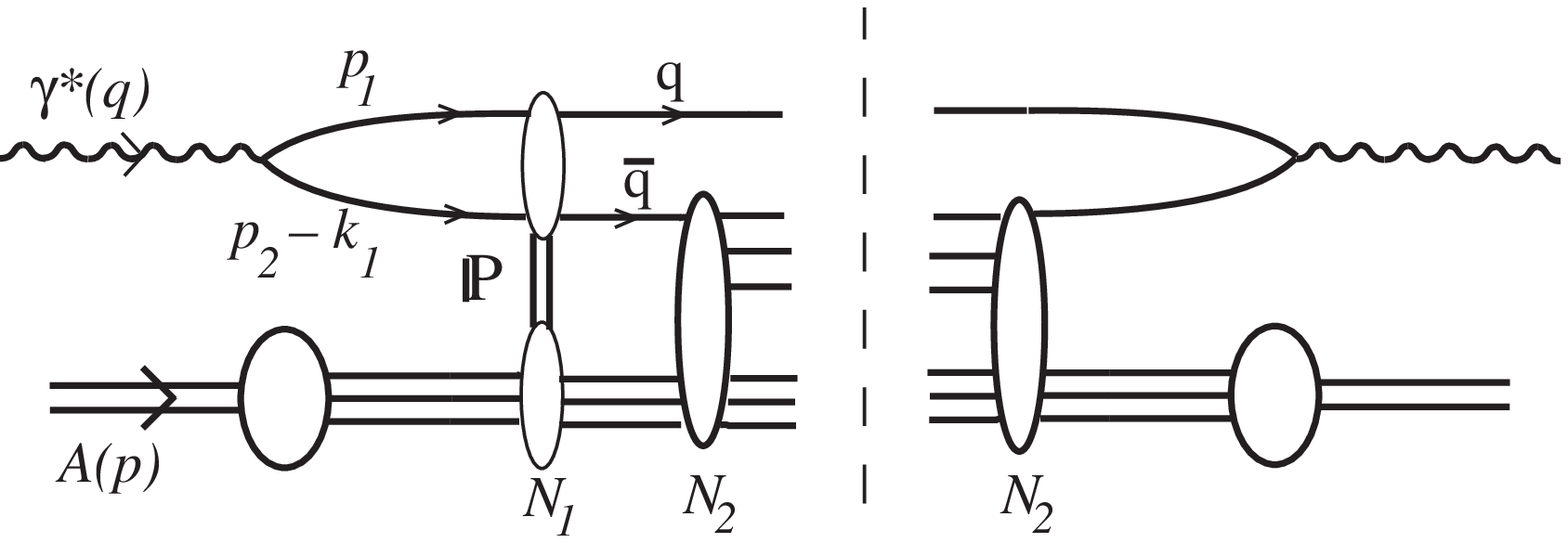}
\caption[*]{Glauber-Gribov shadowing involves interference between
rescattering amplitudes. \label{brodsky2}}
\end{center}
\vspace{.2in}
\begin{center}
\includegraphics[height=2.5in,width=5in]{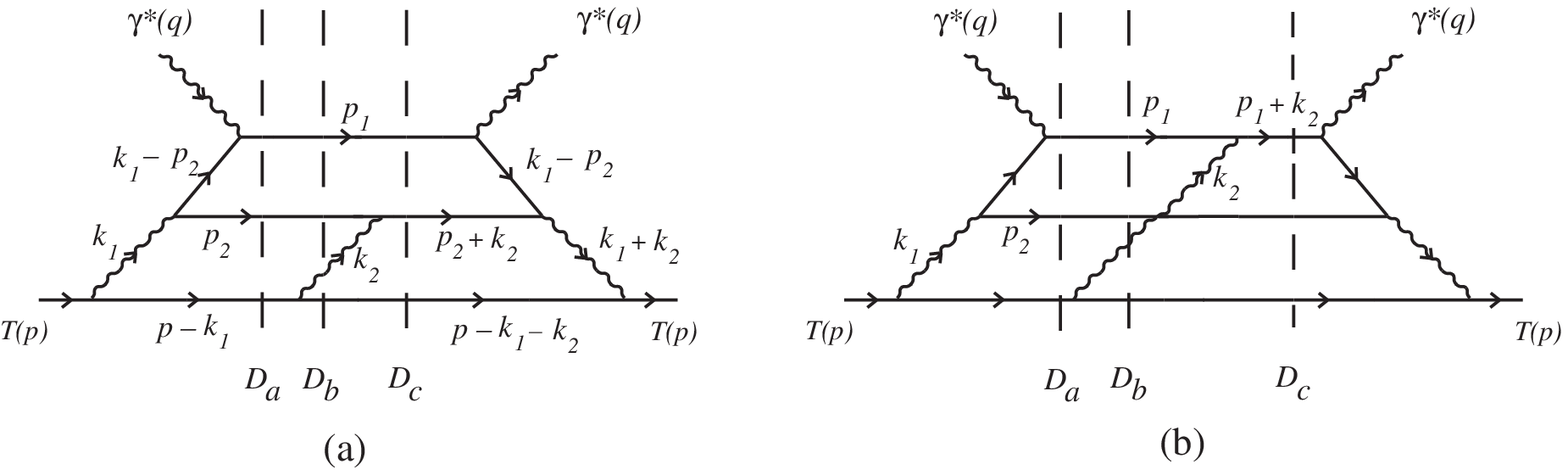}
\caption[*]{Two types of final state interactions.  (a) Scattering
of the antiquark ($p_2$ line), which in the aligned jet kinematics
is part of the target dynamics.  (b) Scattering of the current
quark ($p_1$ line).  For each LC time-ordered diagram, the
potentially on-shell intermediate states---corresponding to the
zeroes of the denominators $D_a, D_b, D_c$---are denoted by dashed
lines.} \label{brodsky1}
\end{center}
\end{figure}

Ever since the earliest days of the parton model, it has been
assumed that the leading-twist structure functions $F_i(x,Q^2)$
measured in deep inelastic lepton scattering are determined by the
{\it probability} distributions of quarks and gluons as determined
by the light-cone (LC) wave functions of the target.  For example,
the quark distribution is \begin{equation} { P}_{\qu/N}(x_B,Q^2)=
\sum_n \int^{k_{iT}^2<Q^2}\left[ \prod_i\, dx_i\, d^2k_{T
i}\right] |\psi_n(x_i,k_{T i})|^2 \sum_{j=q} \delta(x_B-x_j).
\end{equation} The identification of structure functions with the square of
light-cone wave functions is usually made in the LC gauge, $n\cdot A = A^+=0$, where the
path-ordered exponential in the operator product for the forward virtual Compton amplitude
apparently reduces to unity. Thus the deep inelastic lepton scattering cross section (DIS)
appears to be fully determined by the probability distribution of partons in the target.
However, Paul Hoyer, Nils Marchal, Stephane Peigne, Francesco Sannino, and
I~\cite{Brodsky:2002ue} have shown that the leading-twist contribution to DIS is affected
by diffractive rescattering of a quark in the target, a coherent effect which is not
included in the light-cone wave functions, even in light-cone
gauge~\cite{Brodsky:2002ue,Belitsky:2002sm,Collins:2003fm}. The distinction between
structure functions and parton probabilities is already implied by the Glauber-Gribov
picture of nuclear
shadowing~\cite{Stodolsky:1966am,Gribov:1968jf,Brodsky:1969iz,Brodsky:1990qz,Piller:2000wx}.
In this framework shadowing arises from interference between complex rescattering
amplitudes involving on-shell intermediate states, as in Fig.~\ref{brodsky2}.  In contrast,
the wave function of a stable target is strictly real since it does not have on
energy-shell configurations.  A probabilistic interpretation of the DIS cross section is
thus precluded.

It is well-known that in the Feynman and other covariant gauges
one has to evaluate the corrections to the ``handbag" diagram due
to the final-state interactions of the struck quark (the line
carrying momentum $p_1$ in Fig.~\ref{brodsky1}) with the gauge
field of the target.  In light-cone gauge, this effect also
involves rescattering of a spectator quark, the $p_2$ line in
Fig.~\ref{brodsky1}.  The light-cone gauge is singular---in
particular, the gluon propagator \begin{equation}
d_{LC}^{\mu\nu}(k) =
\frac{i}{k^2+\ieps}\left[-g^{\mu\nu}+\frac{n^\mu k^\nu+ k^\mu
n^\nu}{n\cdot k}\right] \label{lcprop} \end{equation} has a pole
at $k^+ = 0$ which requires an analytic prescription.  In
final-state scattering involving on-shell intermediate states, the
exchanged momentum $k^+$ is of \order{1/\nu} in the target rest
frame, which enhances the second term in the propagator.  This
enhancement allows rescattering to contribute at leading twist
even in LC gauge.

The issues involving final-state interactions even occur in the
simple framework of abelian gauge theory with scalar quarks.
Consider a frame with $q^+ < 0$.  We can then distinguish FSI from
ISI using LC time-ordered perturbation theory, LCPTH
\cite{Lepage:1980fj}.  Figure~\ref{brodsky1} illustrates two LCPTH
diagrams which contribute to the forward $\gamma^* T \to \gamma^*
T$ amplitude, where the target $T$ is taken to be a single quark.
In the aligned jet kinematics the virtual photon fluctuates into a
\qu\qb\ pair with limited transverse momentum, and the (struck)
quark takes nearly all the longitudinal momentum of the photon.
The initial \qu\ and \qb\ momenta are denoted $p_1$ and $p_2-k_1$,
respectively.

The calculation of the rescattering effects on DIS in Feynman and
light-cone gauge through three loops is given in detail in
Ref.~\cite{Brodsky:2002ue}.  The result can be resummed and is
most easily expressed in eikonal form in terms of transverse
distances $r_T, R_T$ conjugate to $p_{2T}, k_T$.  The DIS cross
section can be expressed as \begin{equation}
Q^4\frac{d\sigma}{dQ^2\, dx_B} = \frac{\alpha_{\rm
em}}{16\pi^2}\frac{1-y}{y^2} \frac{1}{2M\nu} \int
\frac{dp_2^-}{p_2^-}\,d^2\rvec_T\, d^2\Rvec_T\, |\tilde M|^2
\label{transcross} \end{equation} where \begin{equation} |\tilde{
M}(p_2^-,\rvec_T, \Rvec_T)| = \left|\frac{\sin \left[g^2\,
W(\rvec_T, \Rvec_T)/2\right]}{g^2\, W(\rvec_T, \Rvec_T)/2}
\tilde{A}(p_2^-,\rvec_T, \Rvec_T)\right| \label{Interference}
\end{equation} is the resummed result.  The Born amplitude is
\begin{equation} \tilde
A(p_2^-,\rvec_T, \Rvec_T) = 2eg^2 M Q p_2^-\, V(m_\pl r_T)
W(\rvec_T, \Rvec_T) \label{Atildeexpr} \end{equation} where $
m_\pl^2 = p_2^-Mx_B + m^2 \label{mplus}$ and \begin{equation}
V(m\, r_T) \equiv \int \frac{d^2\pvec_T}{(2\pi)^2}
\frac{e^{i\rvec_T\cdot\pvec_{T}}}{p_T^2+m^2} =
\frac{1}{2\pi}K_0(m\,r_T). \label{Vexpr} \end{equation} The
rescattering effect of the dipole of the \qu\qb~ is controlled by
\begin{equation} W(\rvec_T, \Rvec_T) \equiv \int \frac{d^2\kvec_T}{(2\pi)^2}
\frac{1-e^{i\rvec_T\cdot\kvec_{T}}}{k_T^2}
e^{i\Rvec_T\cdot\kvec_{T}} = \frac{1}{2\pi}
\log\left(\frac{|\Rvec_T+\rvec_T|}{R_T} \right). \label{Wexpr}
\end{equation} The fact that the coefficient of $\tilde A$ in
Eq.~\eq{Interference} is less than unity for all $\rvec_T,
\Rvec_T$ shows that the rescattering corrections reduce the cross
section.  It is the analog of nuclear shadowing in our model.

We have also found the same result for the DIS cross sections in
light-cone gauge.  Three prescriptions for defining the propagator
pole at $k^+ =0$ have been used in the literature:
\begin{equation} \label{prescriptions} \frac{1}{k_i^+} \rightarrow
\left[\frac{1}{k_i^+} \right]_{\eta_i} = \left\{
\begin{array}{cc}
k_i^+\left[(k_i^+ -i\eta_i)(k_i^+ +i\eta_i)\right]^{-1} & ({\rm PV}) \\
\left[k_i^+ -i\eta_i\right]^{-1} & ({\rm K}) \\
\left[k_i^+ -i\eta_i \epsilon(k_i^-)\right]^{-1} & ({\rm ML})
\end{array} \right.
\end{equation} the principal-value (PV), Kovchegov
(K)~\cite{Kovchegov:1997pc}, and Mandelstam-Leibbrandt
(ML)~\cite{Leibbrandt:1987qv} prescriptions.  The `sign function'
is denoted $\epsilon(x)=\Theta(x)-\Theta(-x)$.  With the PV
prescription we have $ I_{\eta} = \int dk_2^+
\left[k_2^+\right]_{\eta_2}^{-1} = 0. $ Since an individual
diagram may contain pole terms $\sim 1/k_i^+$, its value can
depend on the prescription used for light-cone gauge.  However,
the $k_i^+=0$ poles cancel when all diagrams are added.  The net
is thus prescription-independent and agrees with the Feynman gauge
result.  It is interesting to note that the diagrams involving
rescattering of the struck quark $p_1$ do not contribute to the
leading-twist structure functions if we use the Kovchegov
prescription to define the light-cone gauge.  In other
prescriptions for light-cone gauge the rescattering of the struck
quark line $p_1$ leads to an infrared divergent phase factor $\exp
(i\phi)$, where \begin{equation} \phi = g^2 \, \frac{I_{\eta}-1}{4
\pi} \, K_0(\lambda R_{T}) + {{O}}(g^6) \end{equation} where
$\lambda$ is an infrared regulator, and $I_{\eta}= 1$ in the K
prescription. The phase is exactly compensated by an equal and
opposite phase from FSI of line $p_2$.  This irrelevant change of
phase can be understood by the fact that the different
prescriptions are related by a residual gauge transformation
proportional to $\delta(k^+)$ which leaves the light-cone gauge
$A^+ = 0$ condition unaffected.

Diffractive contributions which leave the target intact thus
contribute at leading twist to deep inelastic scattering.  These
contributions do not resolve the quark structure of the target,
and thus they are contributions to structure functions which are
not parton probabilities.  More generally, the rescattering
contributions shadow and modify the observed inelastic
contributions to DIS.

Our analysis in  light-cone gauge resembles the ``covariant parton model" of Landshoff,
Polkinghorne and Short~\cite{Landshoff:1971ff,Brodsky:1973hm} when interpreted in the
target rest frame.  In this description of small $x$ DIS, the virtual photon with positive
$q^+$ first splits into the pair $p_1$ and $p_2$.  The aligned quark $p_1$ has no final
state interactions.  However, the antiquark line $p_2$ can interact in the target with an
effective energy $\widehat s \propto {k_T^2/x}$ while staying close to mass shell.  Thus at
small $x$ and large $\widehat s$, the antiquark $p_2$ line can first multiply scatter in
the target via pomeron and Reggeon exchange, and then it can finally scatter inelastically
or be annihilated.  The DIS cross section can thus be written as an integral of the
$\sigma_{\bar q p \to X}$ cross section over the $p_2$ virtuality.  In this way, the
shadowing of the antiquark in the nucleus $\sigma_{\bar q A \to X}$ cross section yields
the nuclear shadowing of DIS~\cite{Brodsky:1990qz}.  Our analysis, when interpreted in
frames with $q^+ > 0,$ also supports the color dipole description of deep inelastic lepton
scattering at small $x$.  Even in the case of the aligned jet configurations, one can
understand DIS as due to the coherent color gauge interactions of the incoming quark-pair
state of the photon interacting first coherently and finally incoherently in the target.
For further discussion see Refs.~\cite{Brodsky:2001rz,Hoyer:2002fc}.  The same final-state
interactions which produce leading-twist diffraction and shadowing in DIS also lead to
Bjorken-scaling single-spin asymmetries in semi-inclusive deep inelastic, see
Refs.~\cite{Brodsky:2002cx,Collins:2002kn,Efremov:2003tf}.

This analysis has important implications for the interpretation of
the nuclear structure functions measured in deep inelastic lepton
scattering.  Since leading-twist nuclear shadowing is due to the
destructive interference of diffractive processes arising from
final-state interactions (in the $q^+ \le 0$ frame), the physics
of shadowing is not contained in the wave functions of the
isolated target alone.  For example, the light-front wave
functions of stable states computed in light-cone gauge (PV
prescription) are real, and they only sum the interactions within
the bound-state which occur up to the light-front time $\tau=0$
when the current interacts.  Thus the shadowing of nuclear
structure functions is due to the mutual interactions of the
virtual photon and the target, not the nucleus in
isolation~\cite{Brodsky:2002ue}.

\bigskip
\centerline{\large\it  6.  Single-Spin Asymmetries from Final-State
Interactions}
\bigskip

Spin correlations provide a remarkably sensitive window to
hadronic structure and basic mechanisms in QCD.   Among the most
interesting polarization effects are single-spin azimuthal
asymmetries (SSAs) in semi-inclusive deep inelastic scattering,
representing the correlation of the spin of the proton target and
the virtual photon to hadron production plane: $\vec S_p \cdot
\vec q \times \vec p_H$~\cite{Avakian:2002td}.  Such asymmetries
are time-reversal odd, but they can arise in QCD through phase
differences in different spin amplitudes.

The most common explanation of the pion electroproduction
asymmetries in semi-inclusive deep inelastic scattering is that
they are related to the transversity distribution of the quarks in
the hadron $h_{1}$~\cite{Jaffe:1996zw,Boer:2001zw,Boer:2002xc}
convoluted with the transverse momentum dependent fragmentation
function $H^\perp_1$, the Collins function, which gives the
distribution for a transversely polarized quark to fragment into
an unpolarized hadron with non-zero transverse momentum
\cite{Collins93,Barone:2001sp,Ma:2002ns,Goldstein:2002vv,Gamberg:2003ey}.

Recently, an alternative physical mechanism for the azimuthal
asymmetries has been proposed
\cite{Brodsky:2002cx,Collins,Ji:2002aa}.  It was shown that the
QCD final-state interactions (gluon exchange) between the struck
quark and the proton spectators in semi-inclusive deep inelastic
lepton scattering can produce single-spin asymmetries which
survive in the Bjorken limit.  In this case, the fragmentation of
the quark into hadrons is not necessary, and one has a correlation
with the production plane of the quark jet itself $\vec S_p \cdot
\vec q \times \vec p_q.$  This final-state interaction mechanism
provides a physical explanation within QCD of single-spin
asymmetries.  The required matrix element measures the spin-orbit
correlation $\vec S \cdot \vec L$ within the target hadron's wave
function, the same matrix element which produces the anomalous
magnetic moment of the proton, the Pauli form factor, and the
generalized parton distribution $E$ which is measured in deeply
virtual Compton scattering.  Physically, the final-state
interaction phase arises as the infrared-finite difference of QCD
Coulomb phases for hadron wave functions with differing orbital
angular momentum. The final-state interaction effects can be
identified with the gauge link which is present in the
gauge-invariant definition of parton distributions~\cite{Collins}.
When the light-cone gauge is chosen, a transverse gauge link is
required.  Thus in any gauge the parton amplitudes need to be
augmented by an additional eikonal factor incorporating the
final-state interaction and its
phase~\cite{Ji:2002aa,Belitsky:2002sm}.  The net effect is that it
is possible to define transverse momentum dependent parton
distribution functions which contain the effect of the QCD
final-state interactions.    The same final-state interactions are
responsible for the diffractive component to deep inelastic
scattering, and that they play a critical role in nuclear
shadowing phenomena~\cite{Brodsky:2002ue}.

A related analysis also predicts that the initial-state
interactions from gluon exchange between the incoming quark and
the target spectator system lead to leading-twist single-spin
asymmetries in the Drell-Yan process $H_1 H_2^\updownarrow \to
\ell^+ \ell^- X$ \cite{Collins:2002kn,BHS2}.    Initial-state
interactions also lead to a $\cos 2 \phi$ planar correlation in
unpolarized Drell-Yan reactions \cite{Boer:2002ju}.

The single-spin asymmetry (SSA) in semi-inclusive deep inelastic
scattering (SIDIS) $ep^{\updownarrow}\to e'\pi X$, is given by the
correlation ${\vec S}_p\cdot {\vec q}\times {\vec p}_{\pi}$.  For
the electromagnetic interaction, there are two mechanisms for this
SSA: $h_1 H_1^{\perp}$ and $f_{1T}^{\perp}D_1$.  The former was
first studied by Collins~\cite{Collins93}, and the latter by
Sivers~\cite{Sivers}. Hwang, Schmidt and I have calculated
\cite{Brodsky:2002cx} the single-spin Sivers asymmetry in
semi-inclusive electroproduction $\gamma^* p^{\updownarrow} \to H
X$ induced by final-state interactions in a model of a spin-\half
~ proton of mass $M$ with charged spin-\half ~ and spin-0
constituents of mass $m$ and $\lambda$, respectively, as in the
QCD-motivated quark-diquark model of a nucleon.  The basic
electroproduction reaction is then $\gamma^* p \to q (qq)_0$.  In
fact, the asymmetry comes from the interference of two amplitudes
which have different proton spin but couple to the same final
quark spin state, and therefore it involves the interference of
tree and one-loop diagrams with a final-state interaction.  In
this simple model the azimuthal target single-spin asymmetry
$A^{\sin \phi}_{UT}$ is given by
\begin{eqnarray}
A^{\sin \phi}_{UT} &=& {C_F \alpha_s(\mu^2) } \ { \Bigl(\ \Delta\,
M+m\ \Bigr)\ r_{\perp}\over \Big[\ \Bigl( \ \Delta\, M+m\
\Bigr)^2\
+\ {\vec r}_{\perp}^2\ \Big]}\nonumber \\
&\times& \Bigg[\ {\vec r}_{\perp}^2+\Delta
(1-\Delta)(-M^2+{m^2\over\Delta} +{\lambda^2\over 1-\Delta})\
\Bigg] \nonumber\\[1ex] &\times& \ {1\over {\vec r}_{\perp}^2}\
{\rm ln}{{\vec r}_{\perp}^2 +\Delta
(1-\Delta)(-M^2+{m^2\over\Delta}+{\lambda^2\over 1-\Delta})\over
\Delta (1-\Delta)(-M^2+{m^2\over\Delta}+{\lambda^2\over
1-\Delta})}\ . \label{sa2b}
\end{eqnarray}
Here $r_\perp$ is the magnitude of the transverse momentum of the
current quark jet relative to the virtual photon direction, and
$\Delta=x_{Bj}$ is the usual Bjorken variable.  To obtain
(\ref{sa2b}) from Eq. (21) of \cite{Brodsky:2002cx}, we used the
correspondence ${|e_1 e_2|\over 4 \pi} \to C_F \alpha_s(\mu^2)$
and the fact that the sign of the charges $e_1$ and $e_2$ of the
quark and diquark are opposite since they constitute a bound
state.  The result can be tested in jet production using an
observable such as thrust to define the  momentum $q + r$ of the
struck quark.

\begin{figure}[htbp]
\centering
\includegraphics[width=4.9in,height=5.4in]{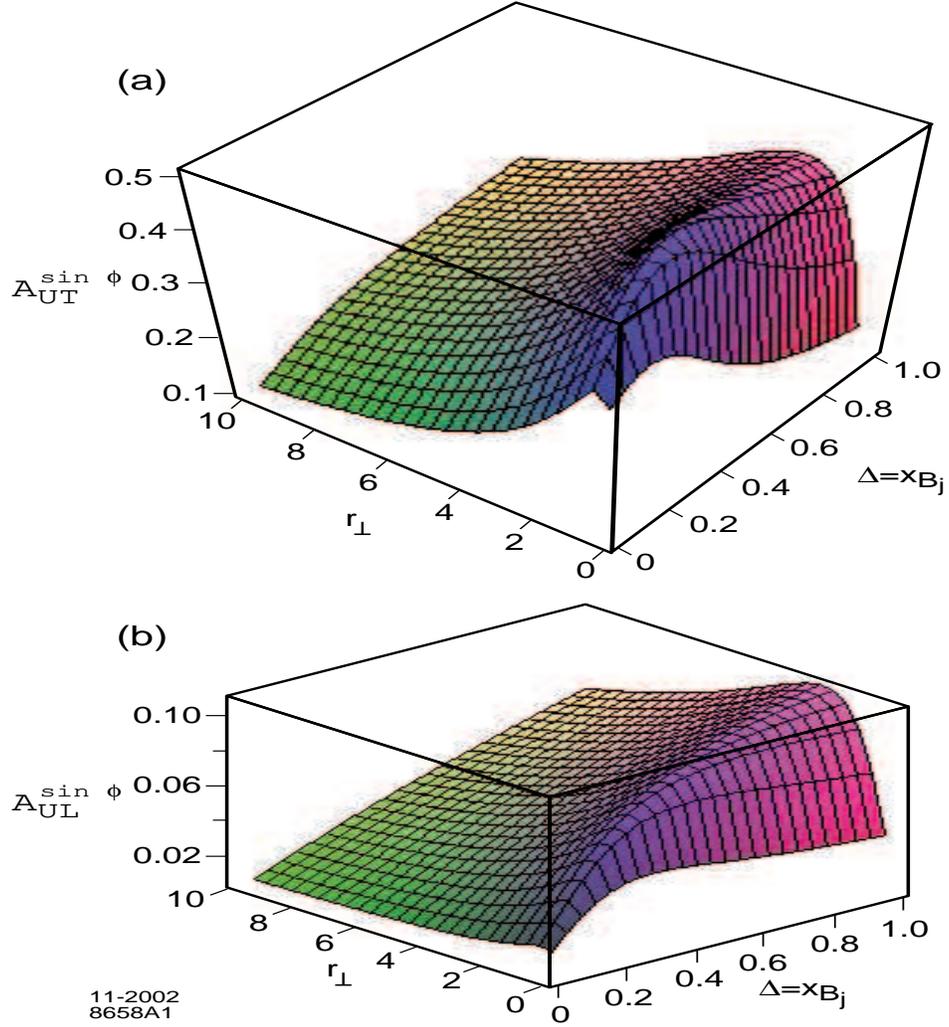}
\caption[*]{\baselineskip=12pt Model predictions for the target
single-spin asymmetry $A^{\sin \phi}_{UT}$ for charged and neutral
current deep inelastic scattering resulting from gluon exchange in
the final state.  Here $r_\perp$ is the magnitude of the
transverse momentum of the outgoing quark relative to the photon
or vector boson direction, and $\Delta = x_{bj}$ is the light-cone
momentum fraction of the struck quark.  The parameters of the model
are given in the text.  In (a) the target polarization is
transverse to the incident lepton direction.  The asymmetry in (b)
$A^{\sin \phi}_{UL} = K A^{\sin \phi}_{UT}$ includes a kinematic
factor $K = {Q\over \nu}\sqrt{1-y}$ for the case where the target
nucleon is polarized along the incident lepton direction.  For
illustration, we have taken $K= 0.26 \sqrt x,$ corresponding to
the kinematics of the HERMES experiment~\cite{Airapetian:1999tv}
with $E_{lab} = 27.6 ~{\rm GeV}$ and $y = 0.5.$} \label{fig:SSA1}
\end{figure}

The predictions of our model for the asymmetry $A^{\sin
\phi}_{UT}$ of the  ${\vec S}_p \cdot \vec q \times \vec p_q$
correlation based on  Eq. ({\ref{sa2b}}) are shown in Fig.
\ref{fig:SSA1} .  As representative parameters we take $\alpha_s =
0.3$, $M =  0.94$ GeV for the proton mass,  $m=0.3$ GeV for the
fermion constituent and $\lambda = 0.8$ GeV for the spin-0
spectator.  The single-spin asymmetry $A^{\sin \phi}_{UT}$ is
shown as a function of $\Delta$ and $r_\perp$ (GeV).  The
asymmetry measured at HERMES~\cite{Airapetian:1999tv}
$A_{UL}^{\sin \phi} = K A^{\sin \phi}_{UT}$ contains a kinematic
factor $K = {Q\over \nu}\sqrt{1-y} = {\sqrt{2Mx\over
E}}{\sqrt{1-y\over y}}$ because the proton is polarized along the
incident electron direction.  The resulting prediction for
$A_{UL}^{\sin \phi}$ is shown in Fig. \ref{fig:SSA1}(b) . Note
that $\vec r = \vec p_q - \vec q$ is the momentum of the current
quark jet relative to the photon momentum. The asymmetry as a
function of the pion momentum $\vec p_\pi$ requires a convolution
with the quark fragmentation function.

Since the same matrix element controls the Pauli form factor, the
contribution of each quark current to the SSA is proportional to
the contribution $\kappa_{q/p}$ of that quark to the proton
target's anomalous magnetic moment $\kappa_p = \sum_q e_q
\kappa_{q/p}$~\cite{Brodsky:2002cx}.
Avakian~\cite{Avakian:2002td} has shown that the data from HERMES
and Jefferson laboratory could be accounted for by the above
analysis.  However, more analysis and measurements especially
azimuthal angular correlations will be needed to unambiguously
separate the transversity and Sivers effect mechanisms.  Note that
the Sivers effect occurs even for jet production; unlike
transversity, hadronization is not required. There is no Sivers
effect in charged current reactions since the $W$ only couples to
left-handed quarks~\cite{Brodsky:2002pr}.

The corresponding single spin asymmetry of the Drell-Yan
processes, such as $\pi p^{\updownarrow}\ ({\rm or}\ p
p^{\updownarrow}) \to \gamma^* X\to \ell^+\ell^- X$, is due to
initial-state interactions.  The simplest way to get the result is
applying crossing symmetry to the SIDIS processes.  The result
that the SSA in the Drell-Yan process is the same as that obtained
in SIDIS, with the appropriate identification of variables, but
with the opposite sign~\cite{Collins,BHS2}.

We can also consider the SSA of $e^+e^-$ annihilation processes
such as $e^+e^-\to \gamma^* \to \pi {\Lambda}^{\updownarrow} X$.
The $\Lambda$ reveals its polarization via its decay $\Lambda \to
p \pi^-$.  The spin of the $\Lambda$ is normal to the decay plane.
Thus we can look for a SSA through the T-odd correlation
$\epsilon_{\mu \nu \rho \sigma} S^\mu_\Lambda p^\nu_\Lambda
q^\rho_{\gamma^*} p^\sigma_{\pi}$.  This is related by crossing to
SIDIS on a $\Lambda$ target.

Measurements from Jefferson Lab~\cite{Avakian:2003pk} also show
significant beam single spin asymmetries in deep inelastic
scattering.  Afanasev and Carlson~\cite{Afanasev:2003ze} have
recently shown that this asymmetry is due to the interference of
longitudinal and transverse photoabsorption amplitudes which have
different phases induced by the final-state interaction between
the struck quark and the target spectators just as in the
calculations of Ref. \cite{Brodsky:2002cx}.  Their results are
consistent with the experimentally observed magnitude of this
effect.  Thus similar FSI mechanisms involving quark orbital
angular momentum appear to be responsible for both target and beam
single-spin asymmetries.

\bigskip
\centerline{\large\it  Acknowledgements}
\bigskip
I wish to thank Professor Ivan Supek and his colleagues at the Rudjer Boskovic Institute
for  hosting an outstanding nuclear and particle physics meeting at Dubrovnik.  I also
thank my collaborators, Carl Carlson, Guy de Teramond, Fred Goldhaber,  Paul Hoyer, Dae
Sung Hwang, Volodya Karmanov, Jungil Lee,  Nils Marchal, Sven Menke, Carlos Merino, Gary
McCartor, Stephane Peigne, Johan Rathsman,   Francesco Sannino, and  Ivan Schmidt, for
their crucial input.  This work was supported by the U.S. Department of Energy, contract
DE--AC03--76SF00515.

\end {document}